# Follow The Money: Exploring the Key Factors Influencing Investment in African Startups

**Khalil Liouane**
Carnegie Mellon University
mliouane@andrew.cmu.edu

**ABSTRACT**

UPDATED—07 May 2023.

The African continent has witnessed a notable surge in entrepreneurial activity, with the number of startups and investments made in the ecosystem growing significantly in recent years. Against this backdrop, this paper presents an in-depth analysis of the critical key factors influencing funding amounts in African startup deals. A comprehensive analysis of 2,521 startup investment deals, spanning from January 2019 to March 2023, was conducted using a combination of statistical and several machine learning techniques.

The results of this study highlight a significant gender diversity gap, the importance of professional experience, and the impact of founders' academic backgrounds. The study reveals that human capital, a diversified sector approach, and cross-border collaboration strategies are crucial for a robust startup ecosystem. Additionally, we identified the potential positive impact of 'Y combinators' for African startups, the implications of exit strategies on deal amounts, and the heterogeneity as well as the incongruity of investment rounds across the continent.

In light of these findings, we propose an assortment of policy recommendations aimed at fostering a propitious milieu for African entrepreneurial ventures, promoting equitable investment distribution, and enhancing cross-border collaboration. By providing a rigorous empirical analysis, this study not only contributes to the existing body of literature but also lays the foundation for future research aimed at promoting investment and catalyzing socio-economic development throughout the African continent.

**INTRODUCTION**

Over the past decade, several regions in Africa have experienced a surge in technology startups, fueled by a young and dynamic population, increasing internet penetration, and widespread mobile adoption [1], creating vibrant and rapidly evolving ecosystems across the continent. Countries such as Nigeria, Kenya, South Africa, and Egypt became hotspots for innovation and entrepreneurship [2], with startups emerging in sectors including fintech, agriculture, health, and renewable energy.

Despite these advancements and the positive growth of both the number of startups and the investment within the continent [2], many African startups grapple with the challenge of securing adequate funding to scale their businesses and achieve their full potential [3]. Funding is a critical aspect of startup success, particularly in the early stages, as it accelerates product development and market entry [4].

This paper seeks to address the research question: What are the key factors influencing the amount of funding raised in startup deals across Africa, and how can these factors inform policy recommendations to optimize investment opportunities for African startups? Investigating this question is vital, as addressing the funding gap could unlock the potential of African startups, enabling them to contribute significantly to solving pressing challenges such as poverty, unemployment, and access to essential services.

By examining the factors that influence funding through qualitative research based on historical data, this study aims to inform governments, development agencies, and private investors in their efforts to develop targeted interventions and create a more enabling environment for entrepreneurship and innovation. This, in turn, could catalyze sustainable economic development across the continent.

In summary, this research paper presents a comprehensive analysis of the factors affecting funding raised by startups in Africa. Through understanding these factors and their implications, the study contributes to the ongoing discourse on fostering a supportive ecosystem for African startups and promoting sustainable economic development across the continent.

**LITERATURE REVIEW:**

The African startup ecosystem has experienced significant growth in recent years, driven by increased investment interest from local and international investors, improvements in technology infrastructure, and a growing pool of talented entrepreneurs. Despite this progress, the African startup ecosystem still faces significant challenges related to the uneven distribution of funding across the continent, with the majority of funding being concentrated in just a few countries and sectors. Therefore, there is a need for more research that focuses on understanding the key factors influencing the amount of funding raised in startup deals across Africa, and how these factors can inform policy recommendations to optimize investment opportunities for African startups.

Several studies have explored the macroeconomic factors that influence venture capital investment in Africa. For example, F. Jaoui, O. Amoussou, and F. H. Kemeze [5] conducted a comprehensive literature review to identify the factors that influence venture capital investment in Africa. They identified several key factors such as GDP growth, the size of the domestic market, and the level of economic freedom. Other factors include the availability of skilled labor, the quality of infrastructure, the strength of legal and regulatory frameworks, and the level of innovation and entrepreneurship in the region. They found that the most significant factors influencing the amount of funding raised in startup deals were the size of the domestic market, the level of economic freedom, and the strength of legal and regulatory frameworks. Similarly, J. Munemo [6] examined the relationship between financial development and entrepreneurship in Africa. They found that financial development has a positive and significant impact on new business density in Africa and that improving all the dimensions of financial development, including depth, access, and efficiency, is much more important for supporting entrepreneurship than focusing on a single dimension such as financial depth. They suggest that policymakers should prioritize raising the quality of financial development in Africa to stimulate entrepreneurship, which can be achieved by enhancing regulatory frameworks, promoting financial literacy, and supporting the development of innovative financial products and services.

Additionally, different reports dived into the trends of startup investment in the continent. For instance, the African Private Capital Association (AVCA), in its annual report on private capital investments, fundraising, and exits in Africa [7], provides valuable insights into private capital investment trends in Africa and sector-specific trends. In its 2023 report, they highlight that financials, consumer discretionary, and industrials continue to attract the lion's share of the total volume of private capital deals, while information technology, healthcare, and utility sectors are on the rise. By the same token, the reports produced by Disrupt Africa [8] as well as by Briter Bridges [2], offer valuable insights into the African startup ecosystem, particularly with regard to funding trends visa a vis sectors, countries, sectors, product types, and stages. The reports confirm the dominance of fintech in the funding landscape, while also identifying other sectors such as cleantech, logistics, mobility, and e-commerce as having significant growth potential. Additionally, both reports discuss the importance of diversity and inclusion within the African startup ecosystem, which is crucial in addressing the barriers faced by underrepresented groups.

While extant literature offers valuable insights into the factors influencing funding in the African startup ecosystem, existing research papers and reports primarily adopt a holistic approach, endeavoring to describe the ecosystem as a whole. Consequently, there remains a lacuna in the literature regarding the specific internal factors that influence funding amounts in startup deals within Africa and the relative importance of these factors. These internal factors may encompass founder characteristics (e.g., gender, education), company attributes (e.g., sector, business model), and investment-related elements (e.g., stage of investment, investor type) and can directly or indirectly impact the deals amounts of startups funding. The aforementioned gap in the literature highlights a need for a more quantitative approach that delves into the impact of different factors, transcending the current holistic understanding of the African startup ecosystem.

**DATA AND METHODS:**

This section of the study provides an overview of the data sources, methodology, and techniques employed to investigate the key factors affecting deal amounts in African startup investments. The data and method section details the process of data collection, cleaning, and preparation, followed by feature grouping, exploratory data analysis, and the implementation of machine learning models to develop a predictive model. The combination of rigorous data handling and state-of-the-art analytical techniques ensures the robustness of the study's findings and enhances their academic credibility.

**Data:**

This study employs a dataset sourced from africathebigdeal.com to systematically investigate the key factors influencing deal amounts in African startup investments and to formulate policy recommendations that bolster the growth of the startup ecosystem on the continent. The dataset compilation adhered to the following methodology:

- Inclusion criteria stipulated that startups must either operate in Africa with their headquarters situated within the continent or possess African founders despite having their headquarters located outside Africa.
- The database exclusively captures deals that have been publicly disclosed or openly shared by investors or founders themselves.
- Deal size limitations dictate the inclusion of transactions amounting to a minimum of

+$100,000 for 2023, 2022, and 2021; +$500,000 for 2020; and +$1,000,000 for 2019.

The principal dataset, serving as the primary focus of this investigation, comprises 2,521 startup deals, encompassing 34 attributes, including the specific deal amount. Simultaneously, the secondary dataset comprises information regarding 1,792 investors who engaged in a minimum of one investment in African startups.

**Data Preprocessing:**

The process of preparing and cleaning the data involved several sequential steps, as outlined below:

1. Scrutinizing both datasets for any missing or erroneous data points and addressing them accordingly by imputing or removing the data points as appropriate, as well as rechecking the integrity of the data from media releases.
2. Integrating the two datasets by merging them based on the investor's name, which resulted in a comprehensive dataset containing both startup and investor information.
3. Transformation of categorical variables such as sector and deal type into binary variables or dummy variables, to be utilized in the analysis.
4. Standardizing numerical variables such as amount raised and valuation to facilitate comparability across diverse units of measurement.

**Feature Grouping:**

To further understand the implications of the different key factors, the features extracted from the primary and secondary datasets were organized into three distinct categories, as outlined below:

- Founding team features (F): This group includes attributes related to the startup's founding team, such as the number of founders, gender-mix, presence of a woman co-founder or CEO, the CEO's university, country and continent of the university, graduation year, and the years elapsed between graduation and the startup's launch.
- Company-related features (C): This category encompasses variables associated with the startup itself, such as the name, website, country, and region of operation, launch date, sector, number of employees, and a brief description of the business.
- Investment-related features (I): This group consists of variables related to the investment deals, including the deal year and date, type of investment, valuation, exit status, investor details, and whether the startup is a Y Combinator alumnus.

Through diligent feature grouping, the study ensured that the variables are organized in a coherent manner, ultimately enhancing the clarity and interpretability of the analysis.

**Exploratory Data Analysis (EDA):**

In this study, Exploratory Data Analysis (EDA) was conducted to examine the dataset and identify key factors that affect deal amounts in African startup investments. A critical aspect of EDA was assessing correlations between variables. Using Pearson's correlation coefficient, we measured the linear association between the dependent variable (deal amount) and the independent variables (founding team, company, and investment-related features). In order to better investigate the correlation between the features, we used the three feature groups discussed earlier in 5 combinations: F, C, I, F+C, and F+C+I. This approach aimed to uncover complex relationships between variables and better understand the importance of each feature.

**Models:**

Using the same combinations of feature groups discussed in the EDA section, four machine learning algorithms were employed: Linear Regression (LR), Support Vector Regression (SVR), Random Forest (RF), and Distributed Gradient Boosting (DGB). Each model was trained and tested using cross-validation. To evaluate the prediction models, the Mean Squared Error (MSE) metric was employed.

During the cross-validation process, the performance of each model was assessed by averaging the MSE values obtained from each fold. The comparison of these averaged MSE values facilitated the selection of the most accurate and reliable algorithm for predicting funding amounts in African startups. The chosen model's performance, along with insights gained from the EDA process, served as the foundation for policy recommendations aimed at supporting the growth of the African startup ecosystem.

**RESULTS:**

This section of the study provides an overview of the results of the EDA, the correlation between the deal amount and the other features, and the model results based on different grouping of features.

**Deals vs Countries:**

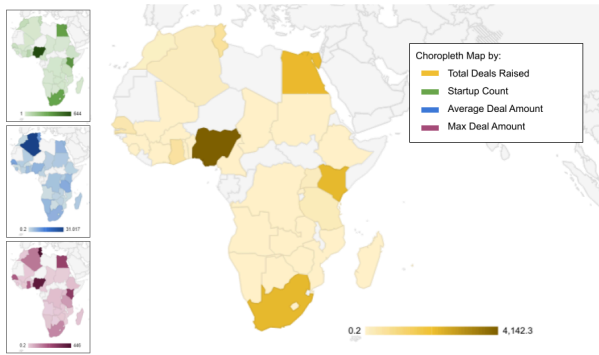

**Figure 1. Chorpleth maps for Total Deals Raised, Startup Count, Average Deal Amount, and Maximum Deal Amount by country.**

The chorpleth maps highlight the varying levels of fundraising success and the concentration of startups in specific countries and illustrate the relationship between the total amount raised (yellow), the number of startups(green), average deal amount (blue), and the maximum deal amount (purple) for each of the analyzed countries. The graph provides a clear visual representation of the disparities in fundraising across different African countries. Nigeria, Kenya, South Africa, and Egypt emerge as the top technology hubs, with the highest total amount raised and a significant number of startups. However, it is also noteworthy that countries such as Algeria and Tunisia and Algeria have higher average and maximum deal amounts respectively, which may indicate a different investment landscape in these countries.

**Deals and gender:**

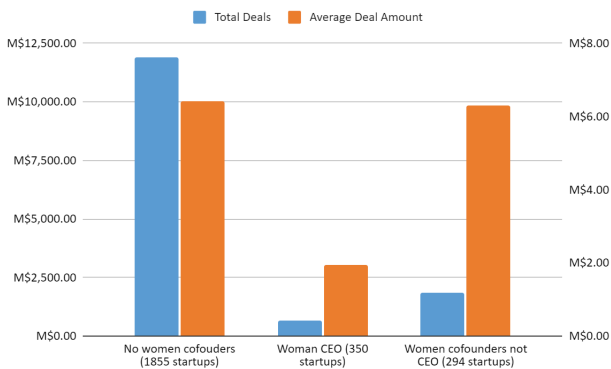

**Figure 2. Total deal amount and average deal amount in M$ based on the gender of founders and CEO**

The bar plot, presents a comparison of the number of startups, total deals, and average deal amounts across three categories: startups with no women co-founders, startups with a woman CEO, and startups with women co-founders but not a woman CEO. The plot vividly illustrates the substantial gender imbalance in the African startup ecosystem. The startups with no women co-founders constitute the majority, accounting for 1,855 startups, and raising a total of M$11,891.10 with an average deal amount of M$6.41. In contrast, startups with a woman CEO represent a relatively smaller segment, comprising 350 startups, raising M$678.70, and having an average deal amount of M$1.94. Startups with women co-founders but not a woman CEO encompass 294 startups, raising M$1,855.60, and achieving an average deal amount of M$6.31.

**Deals and type of investment:**

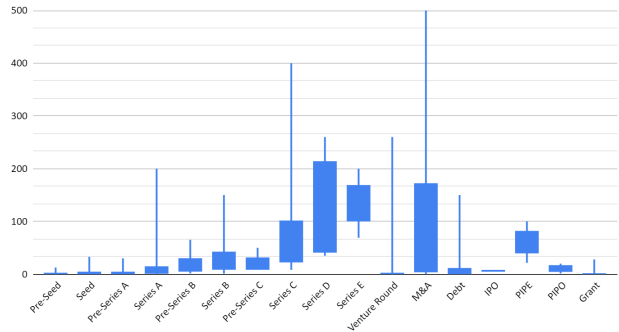

**Figure 3. Box plot of the amount raised in M$ based on the type of investment.** [1]

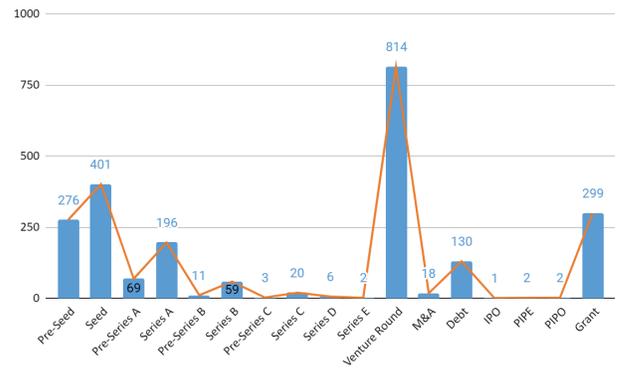

**Figure 4. The number of deals across stages**

The box plot (figure 3) offers a comprehensive visualization of the distribution of deal amounts across various investment types in the African startup ecosystem, revealing considerable variation among them. For instance, Pre-Seed investments exhibit a narrow range, with a minimum of M$0.1, Q1 at M$0.2, Q3 at M$1.0, and a maximum of M$12.6. Conversely, Series C investments display a broader range, with a minimum of M$8.2, Q1 at M$24.0, Q3 at M$100.0, and a maximum of M$400.0. The increasing deal amounts across investment rounds signify the escalating capital requirements of startups at more advanced stages. The box plot also underscores the inconsistency in amounts per investment type, particularly for the primary investment rounds (Pre-Seed, Seed, Series A, Series B, Series C, and Series E), suggesting a lack of clear term definitions within the ecosystem. This implies that the investment landscape may be influenced by various factors beyond the investment stage.

---

[1] "Venture round" is a term that refers to an investment round that has been announced by the startup and the respective investors without specifying the proper label (pre-seed, pre-series A, pre-series B, pre-series C) at the time of the transaction. In some cases, the venture round might be changed later on to the proper label. However, for the sake of this study, we focus on how the transaction was announced.

In contrast, the bar plot (figure 4) depicts the frequency distribution of investment types, demonstrating a general decline in investment frequency as rounds progress, with fewer instances in later stages such as Series D (6), Series E (2), and IPO (1). However, Venture Round investments exhibit a significantly higher frequency (814), likely because both startups and investors regard this step as a precursor to the main fundraising rounds. Moreover, other funding sources such as Grants (299) and Debt (130) display notable occurrences.

The relatively low frequency of M&A deals (18), IPOs (1), and PIPE (2) indicates a lack of development in Africa's entrepreneurial ecosystem. This observation, combined with the findings from the box plot, highlights the need for further research and policy interventions to support the growth and sustainability of the African startup ecosystem, as well as to address inconsistencies and gaps in the current investment landscape.

**Deals and sector implications:**

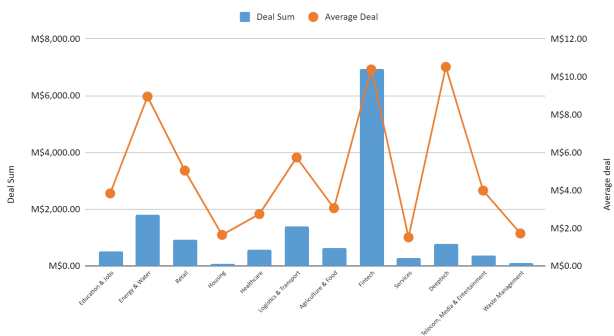

**Figure 5. Total deal amount and average deal amount per sector**

The graph presents a combined visualization of the deal sum and average deal amount by sector within the African startup ecosystem, using a bar plot and line plot, respectively. The bar plot illustrates the total deal sum for each sector, with Fintech emerging as the dominant sector with a total of M$6,947.90 in investments. Energy & Water (M$1,808.70) and Logistics & Transport (M$1,381.90) follow as the second and third highest sectors in terms of the deal sum. In contrast, the Housing (M$77.30) and Waste Management (M$99.80) sectors exhibit the lowest total investments, indicating that these sectors may be less attractive to investors or may lack the necessary infrastructure and support for growth. Whereas, the line plot represents the average deal amount per sector, with Deeptech (M$10.52) and Fintech (M$10.39) sectors showing the highest average deal sizes. This suggests that these sectors typically attract larger investments, possibly due to their potential for disruptive innovation and scalability. On the other hand, the Services (M$1.52) and Waste Management (M$1.72) sectors display the lowest average deal amounts, indicating that investments in these areas may be relatively modest or focused on smaller-scale projects.

**Deals and Y combinator implications:**

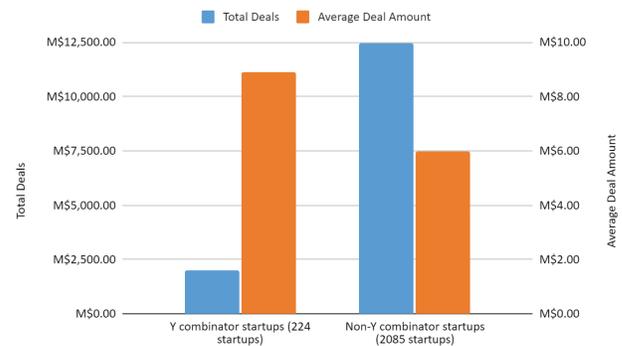

**Figure 6. Total deal amount and average deal amount based on the involvement of the startup in Y Combinator**

The graph provides a visual comparison of the total deals, average deal amount, and count of startups for two distinct categories within the African startup ecosystem: Y Combinator-backed startups (224 startups) and non-Y Combinator-backed startups (2,085 startups). This comparison allows for an assessment of the potential impact of Y Combinator affiliation on startup investments. The graph illustrates that Y Combinator-backed startups have raised a total of M$1,992.00 in deals, while non-Y Combinator-backed startups have raised a considerably higher sum of M$12,433.40. However, it is essential to consider the difference in the number of startups in each category. There are 224 startups in the Y Combinator category, while the non-Y Combinator category comprises 2,085 startups.

The average deal amount, which accounts for the varying number of startups in each category, reveals a different pattern. Y Combinator-backed startups exhibit a higher average deal amount (M$8.89) compared to their non-Y Combinator counterparts (M$5.96). This suggests that startups affiliated with Y Combinator tend to attract larger investments, possibly due to the accelerator's reputation, resources, and mentorship.

**Correlation Matrices:**

This sub sub-section presents the results of correlation matrices to explore the relationships between various features and their impact on deal amounts in the African startup ecosystem. The analysis is conducted by examining different groupings of features and identifying those that exhibit high positive or high negative correlations with deal amounts. Table 1 shows the top two features with high positive correlations while Table 2, shows the top two features with high negative correlations for each group. These correlations provide insights into the factors that may influence deal amounts and inform potential strategies for entrepreneurs, investors, and policymakers to foster growth in the African startup ecosystem.

For instance, the analysis reveals that the number of employees and the exit type have high positive correlations with deal amounts, while the CEO's graduation year and the presence of a woman CEO show high negative correlations. This information can guide stakeholders in understanding the dynamics of the startup ecosystem and tailoring their strategies accordingly.

By examining the correlation matrices, we aim to deepen our understanding of the African startup landscape and identify trends that may influence investment decisions, startup growth, and overall ecosystem development.

| Groups used | Features with High Positive Correlation | Correlation coefficient |
|---|---|---|
| F | # of Founders | 0.10 |
|   | CEO University/School - Continent | 0.09 |
| C | # of employees | 0.20 |
|   | Website | 0.03 |
| I | Exit | 0.33 |
|   | Type | 0.001 |
| F+C | # of employees | 0.24 |
|   | # of Founders | 0.10 |
| F+C+I | **Exit** | **0.35** |
|   | # of employees | 0.24 |

Table 1. The two features that are highly positively correlated with the deal amount by group feature

| Groups used | Features with High Negative Correlation | Correlation coefficient |
|---|---|---|
| F | CEO Graduation year | -0.13 |
|   | Woman CEO | -0.06 |
| C | Sector | -0.05 |
|   | Country | -0.02 |
| I | 1st $M ? | -0.17 |
|   | Deal Year | -0.03 |
| F+C* | Country | -0.01 |
| F+C+I | 1st $M ? | -0.18 |
|   | **Launch** | **-0.20** |

Table 2. The two features that are highly negatively correlated with the deal amount by group feature (* only one negative correlated feature).

**Model Performance:**

| Features used | Algorithm | MSE |
|---|---|---|
| F | **LR** | **19.89** |
|   | SVR | 36.93 |
|   | RF | 20.22 |
|   | DGB | 20.49 |
| C | **LR** | **24.27** |
|   | SVR | 50.31 |
|   | RF | 24.37 |
|   | DGB | 24.72 |
| I | LR | 23.65 |
|   | SVR | 41.75 |
|   | **RF** | **19.16** |
|   | DGB | 19.73 |
| F+C | **LR** | **19.52** |
|   | SVR | 43.50 |
|   | RF | 20.26 |
|   | DGB | 20.20 |
| F+C+I | LR | 19.02 |
|   | SVR | 32.24 |
|   | **RF** | **16.80** |
|   | DGB | 16.86 |

Table 3. Model performance evaluation using MSEs for the features groups and the algorithms used.

The table above rigorously assessed the efficacy of various machine learning algorithms used in predicting deal amounts in African startup investments, contingent upon specific groups of features. The analysis unveiled disparate model performance contingent on the features under consideration, underscoring the necessity of tailoring model selection to the pertinent features.

Linear Regression, adept at estimating coefficients of predictor variables to fit a linear equation to the data, exhibited superior performance for features F, C, and F+C, as evidenced by MSE values ranging from 19.524 to 24.272. Conversely, the Random Forest and Gradient Boosting models demonstrated commendable performance across all features, particularly for F+C+I, suggesting their proficiency in capturing intricate relationships and interactions among features. In comparison, Support Vector Regression (SVR) displayed elevated MSE values for all features, indicative of suboptimal performance relative to the other models.

The Random Forest model displayed optimal performance for the I and F+C+I features, implying its suitability as the most appropriate model for predicting deal amounts in African startup investments. The confluence of founder features (F), company-related features (C), and investment/investor-related features (I) emerged as critical predictor of deal amounts, as evidenced by the lowest MSE values across all models. The consistent performance of the combination of F, C, and I (F+C+I) accentuates its significant impact on the prediction of deal amounts in African startup investments.

The consistent performance of the F+C+I feature group with the 4 algorithm used, highlight the substantial influence of this group on predicting deal amounts in African startup investments. Thus, it is imperative to investigate the correlation coefficients between the deal amount and other features within the F+C+I group. A deeper understanding of these correlations will offer valuable insights into the relative importance of each feature, enabling us to make informed decisions and adopt targeted strategies to foster the growth in the discussion section.

| Feature | Correlation coefficient |
|---|---|
| Exit[2] | 0.349 |
| # of employees | 0.244 |
| # of Founders | 0.102 |

---

[2] "Exit" was studied as a factor that can impact the transaction amount.

| | |
|---|---|
| CEO - University/School - Continent | 0.093 |
| Founders gender mix | 0.080 |
| Founder 3 | 0.075 |
| CEO - University/School - Country | 0.068 |
| Founder 2 | 0.047 |
| Founder 4 | 0.046 |
| Years between graduation and startup launch | 0.028 |
| Website | 0.028 |
| CEO - University/School | 0.026 |
| Start-up name | 0.023 |
| Description | 0.017 |
| Founder 1 (CEO) | 0.015 |
| Deal Year | 0.012 |
| Country | -0.009 |
| Y combinator | -0.010 |
| Type | -0.015 |
| Region | -0.026 |
| Woman co-founder | -0.038 |
| Women only founders | -0.052 |
| Sector | -0.058 |
| Woman CEO | -0.059 |
| CEO Graduation year | -0.128 |
| 1st $M ? | -0.177 |
| Launch | -0.201 |

**Table 4. Correlation coeffetients between deal amount and the different features within F+C+I group. (From High to Low)**

**DISCUSSION:**

The Discussion section aims to contextualize and synthesize the findings of our research, examining their implications for the African startup ecosystem and highlighting potential avenues for further inquiry. In this study, we have unveiled critical insights regarding the impact of various factors on deal amounts in African startup investments, ranging from gender diversity and founder's academic background to the role of incubators, exit strategies, entrepreneurial experience, and sector preferences. By dissecting these factors and understanding their influence on investment outcomes, we seek to contribute to the ongoing discourse on fostering a sustainable and thriving African startup ecosystem. The following subsections will delve into the implications of each finding, shedding light on potential areas for future research and policy interventions to address the challenges and harness the opportunities identified in this study.

**Founder Related Factors:**

*Women Impact on Investment:*

Our analysis revealed a significant gender imbalance in the African startup ecosystem, with 75.2% of startups lacking female co-founders and a mere 10% having female CEOs. Interestingly, the presence of female-only founders, women co-founders, and women CEOs displayed a slight negative correlation with the deal amount. This finding raises concerns about the perceived value of gender diversity within the ecosystem and highlights the existence of a considerable gender diversity gap that warrants further investigation and rectification. Furthermore, when comparing the average deal amounts raised by startups with women CEOs and those with women co-founders, our findings suggest that the increasing policies supporting women entrepreneurs in African countries [9] may inadvertently lead to negative consequences, such as tokenism. To address these issues, policymakers should consider devising targeted interventions that not only foster female entrepreneurship but also promote a genuine appreciation for gender diversity in the startup ecosystem. Additionally, efforts should be made to ensure that funding decisions are based on merit and the potential for success, rather than merely fulfilling diversity quotas. Future research should investigate the relationship between these policies and the observed discrepancy in deal amounts, to better understand the underlying dynamics and inform more effective policy interventions for promoting genuine gender equality and inclusion.

*CEO graduation year:*

Our investigation revealed a negative correlation between this variable and the transaction size, indicating that entrepreneurs with greater experience are potentially more adept at securing funding. This finding may be attributed to the difference between necessity and opportunity-driven entrepreneurship [10]. The implications of this observation encompass a wider inquiry into the significance of entrepreneurial experience in the African startup arena and the probable trade-offs between experience and innovation. Policymakers and investors should endeavor to devise mechanisms that foster the growth of both experienced entrepreneurs and young innovators, recognizing that different types of entrepreneurs may contribute divergently to the ecosystem's advancement and fortitude.

*CEO university impact:*

The results of our study demonstrated a higher probability of African startup fundraising by founders with North American academic backgrounds. Nonetheless, it is vital to emphasize that this correlation does not invariably guarantee startup triumph. This discovery raises inquiries regarding the impact of global exposure and the potential effects of academic history on investment possibilities in Africa. Further investigation is warranted to elucidate whether this tendency is propelled by networking prospects, perceived reputation, or other education-related factors, and how these elements can be harnessed to enhance the African startup ecosystem.

**Company Related Factors:**

*Human capital importance:*

Our analysis revealed that both features: the number of founders and the number of employees are highly positively correlated with the deal amount. This finding highlights the significance of human capital in securing funding for African startups, as investors seem to value the presence of a competent and diverse team in their investment decisions. Startups with more founders and employees may be perceived to have a higher potential for innovation, operational efficiency, and scalability, which could enhance their attractiveness to investors [11]. Hence, it is recommended that policymakers and investors prioritize policies that foster human capital development in startups, including training programs for entrepreneurs and their employees, and policies that encourage diverse founding teams.

*Sector impact:*

The fintech industry emerged as the most prominent sector; however, it was not a strong predictor of the transaction size. The top five sectors align with the challenges and opportunities for innovation and growth on the continent. This outcome indicates that while some sectors may garner more investment and attention, other sectors may possess the untapped potential for expansion and progress. A more diversified and equitable approach to investments in specific sectors may contribute to fostering a more resilient and comprehensive startup ecosystem in Africa. Additional research should delve into the constituents that propel the attractiveness of specific sectors and assess how such insights can guide investment strategies and policy interventions.

*Country Impact:*

Our analysis revealed negligible correlation between country and deal amounts in African startup investments, despite differences in various country-level indices, such as GDP per capita, corruption indices, ease of doing business, and economic development. Additionally, the analysis revealed a strong correspondence between the number of startups and the total sum of deals in each country, with the top four countries—Nigeria, Kenya, South Africa, and Egypt—emerging as prominent technology hubs in Africa. Interestingly, when examining the average deal amount per country, Algeria[3] emerged as the top country, while Tunisia[4] took the lead in terms of the maximum deal amount. These findings prompt questions about the factors that may contribute to the observed discrepancies in deal amounts among different countries and the implications for the broader African startup ecosystem. One plausible explanation for the higher average and maximum deal amounts in Algeria and Tunisia may be their proximity to Europe, which could facilitate access to European markets, investors, and resources. Moreover, while the 4 African tech-hubs seem to foster more startup investments overall, they may not be providing the necessary infrastructure for these startups to scale significantly.

Based on these findings, policymakers should consider implementing targeted strategies to promote cross-border collaboration and facilitate access to international markets, investors, and resources. By fostering such connections, African countries can better leverage their geographical advantages and enhance the growth and sustainability of their respective startup ecosystems. Future research should investigate the underlying factors of this lack of correlation and examine the drivers of success in countries with higher average or maximum deal amounts. Identifying these drivers can help stakeholders develop targeted strategies to support the growth and sustainability of Africa's startup ecosystem.

**Investment Related Factors:**

*Y Combinator's role:*

Albeit our findings did not reveal a significant influence of Y Combinator on the transaction size, the US accelerator asserts that its supported startups raise 2.5 times higher valuation after participating in their program and that 4% of their companies became unicorns[5]. This disparity necessitates further exploration to comprehend the function of incubators and accelerators in the African startup milieu. A productive avenue for research would entail an examination of the constituents that contribute to the success of Y Combinator-sponsored startups and an inquiry into how such insights can inform the development and execution of incubator and accelerator programs in Africa.

*Exit impact:*

Exit strategies play a crucial role in startup investment, as they provide a mechanism for early-stage investors (notable VC firms) to realize their returns on investment. Our analysis revealed a strong positive correlation between the potential for a lucrative exit and the transaction size. However, our analysis showed that there is a dearth of successful exit deals (less than 1%), creating challenges for startups seeking funding and investors seeking returns. This may be due to a variety of factors, including a lack of established companies with the resources to acquire startups, limited public market access, and limited infrastructure for initial public offerings (IPOs) or mergers and acquisitions (M&A). Therefore, to foster a more sustainable startup ecosystem in Africa, policymakers and investors should consider implementing strategies that facilitate the creation of a more conducive environment for successful exits. This may include supporting the growth of large companies and the development of public markets, as well as creating more avenues for mergers and acquisitions or other exit mechanisms. By promoting a more robust exit ecosystem, stakeholders can incentivize investors to fund more startups and provide entrepreneurs with the resources they need to grow their businesses.

*Type of investment impact:*

During the investigation of the relationship between the type of investment and deal amounts in African startup investments, the present study found limited evidence of a significant impact of investment type on the deal amounts. Surprisingly, the study observed substantial inconsistency in the amounts per type of investment, particularly for the primary types of investments, including Pre-seed, Seed, Series A, Series B, Series C, and Series E. Notably, these investments are typically announced as "venture rounds" rather than being properly labeled with their respective investment types.

This observation raises pertinent questions regarding the potential consequences of such ambiguity on investment decisions and the overall health of the ecosystem. The lack of clarity and consistency in the classification of investment stages may lead to discrepancies in investor expectations, founder valuations, and the allocation of resources. Consequently, this may result in inefficient deployment of capital and suboptimal growth for startups in the African continent. Future research should investigate the factors contributing to this inconsistency and explore potential solutions to standardize the classification of investment stages in the African startup ecosystem. By promoting a more consistent and transparent approach to investment categorization, stakeholders can facilitate informed decision-making, foster investor confidence, and support the development of a more robust and sustainable startup landscape in Africa.

---

[3] The number of startup deals was low (6), and one startup, Yassir (https://yassir.com/en/home/), increased the overall average of the country since it raised 150M$ in November 2022.

[4] The maximum deal was made by the Tunisian AI startup, Instadeep (https://www.instadeep.com/), that got acquired for 684M$ by German biotech company BioNTech SE. (https://disrupt-africa.com/2023/01/11/tunisia-founded-ai-startup-instadeep-acquired-in-684m-deal/)

[5] Information provided through the website of Y combinator available here https://www.ycombinator.com/

**CONCLUSION:**

In conclusion, this study has provided valuable insights into the key factors influencing the amount of funding raised in startup deals across Africa. Our findings highlight the significance of various factors on deal amounts in African startup investments, including gender diversity, founder's academic background, the role of incubators, exit strategies, entrepreneurial experience, and sector preferences. These factors have crucial implications for the African startup ecosystem and can inform policy recommendations to optimize investment opportunities for African startups.

The study contributes to the ongoing discourse on fostering a supportive ecosystem for African startups and promoting sustainable economic development across the continent. To harness the potential of African startups in addressing pressing challenges such as poverty, unemployment, and access to essential services, targeted interventions and an enabling environment for entrepreneurship and innovation must be created. Policymakers, development agencies, and private investors should leverage the findings of this study to devise strategies that address the identified challenges and capitalize on the opportunities presented by the African startup landscape.

Moreover, the research underlines the need for further investigation into the underlying factors behind the observed correlations and the drivers of success in countries with higher average or maximum deal amounts. Future research should also explore the role of incubators and accelerators, the need for standardization of investment stages, and potential avenues for cross-border collaboration and access to international markets. By promoting a more consistent and transparent approach to investment categorization, fostering investor confidence, and supporting the development of a more robust and sustainable startup landscape in Africa, stakeholders can contribute to the long-term growth and prosperity of the continent.

**APPENDIX:**

**Model Interpretability:**

The interpretability of a model refers to the ease with which its decision-making process and inner workings can be understood. In this analysis, the Linear Regression model emerged as the most interpretable model due to its ability to fit a linear equation to the data, estimating the coefficients of the predictor variables. The coefficients directly represent the effect of each predictor variable on the deal amount, thus enabling easy comprehension of the relationships between the features and the response variable. Conversely, the Random Forest and Gradient Boosting models are more complex, as they are ensemble techniques that combine multiple decision trees to make predictions. While it is possible to extract feature importance from these models, understanding the interactions and relationships among features may be more challenging. Additionally, Support Vector Regression (SVR) is also a relatively complex model, as it maps the input data into a higher-dimensional space and finds an optimal hyperplane that best separates the data points. Therefore, it is crucial to consider the level of interpretability required when selecting a model for a given analysis.

**Model Comparison:**

Our analysis revealed that the Linear Regression model outperformed the other models for features F, C, and F+C, but underperformed for the F+C+I feature set. Meanwhile, both the Random Forest and Gradient Boosting models showed relatively good performance across all feature sets, with the Random Forest model outperforming the other models for the F+C+I feature set. The SVR model demonstrated the poorest performance across all feature sets, as indicated by its high MSE values. Regarding interpretability, the Linear Regression model is the most straightforward to understand, while the Random Forest, Gradient Boosting, and SVR models are more complex. However, due to its ability to capture non-linear relationships and interactions and its relatively good performance across all feature sets, the Random Forest model appears to be the most suitable model for this analysis.

**Model Assumptions and Limitations:**

Linear Regression is founded on the assumption of a linear relationship between the response variable and the predictor variables. However, this premise may not hold true for all datasets. Moreover, it presumes normally distributed residuals and predictor variables that are not highly correlated (multicollinearity). The violation of these assumptions may result in coefficient estimates and unreliable predictions. Random Forest and Gradient Boosting models, on the other hand, are less sensitive to linearity and normality assumptions. Nonetheless, they can suffer from overfitting when the number of trees is too large, or the depth of each tree is too high. In addition, they can be computationally demanding, particularly when the dataset is large, or a large number of hyperparameters needs optimization. SVR assumes that data can be separated by a hyperplane in a higher-dimensional space, which may not always be the case. Furthermore, the performance of SVR is sensitive to the choice of kernel function and hyperparameters.

In conclusion, although each model has its unique strengths and limitations, the Random Forest model seems the most fitting for this particular analysis. Nevertheless, it is crucial to consider the assumptions and constraints of each model while interpreting the outcomes and applying such a complex problem of predicting deal amounts in the African startup investment context based on different features.